\documentclass[12pt]{article}
\usepackage{graphicx}

\begin{document}
\title{Inverse cascade in percolation model: 
hierarchical description of time-dependent 
scaling}
\author{Ilya Zaliapin, Henry Wong, Andrei Gabrielov}

\maketitle
\tableofcontents
\newpage

\begin{abstract}
The dynamics of a 2{\it D} site percolation model
on a square lattice is studied using the
hierarchical approach introduced 
by Gabrielov et al., {\it Phys. Rev. E}, {\bf 60}, 
5293-5300, 1999. 
The key elements of the approach are the tree
representation of clusters and their coalescence, 
and the Horton-Strahler scheme for cluster ranking.
Accordingly, the evolution of percolation
model is considered as a hierarchical inverse 
cascade of cluster aggregation.
A three-exponent time-dependent scaling for the cluster 
rank distribution is derived using the Tokunaga
branching constraint and classical results on
percolation in terms of cluster masses.
Deviations from the pure scaling are described.
An empirical constraint on the dynamics
of a rank population is reported based on
numerical simulations.
\end{abstract}

\section{Introduction}
Percolation model is probably the simplest 
and best studied system that experiences (geometrical) 
phase transition of the second kind \cite{SA}.
It is widely used as a toy model for 
spatially distributed stochastic processes, 
such as diffusion in disordered media, 
forest fires, gelation, semiconduction, etc.
\cite{Sornette04,SA}.
Importantly for our study, percolation model 
presents a transparent mechanism of the 
process of hierarchical aggregation
(coagulation).
This process has been actively employed for describing 
the essential properties of material fracture and 
earthquake nucleation
\cite{BSLA97,BSL97,GKZN00,NS90,NG91,NTG95,SB01,ZKG03},
starting from the pioneering works of 
Allegre et al. \cite{ALP82} and 
Newman and Knopoff \cite{NK82,NK83,KN83}.
In this paper we describe the evolution of percolation 
model in terms of an inverse cascade of 
hierarchical cluster aggregation.

An early idea of hierarchical aggregation was 
introduced by Newman and Knopoff in the ``crack-fusion'' 
model for repetitive cycles of large earthquakes
\cite{NK82,NK83,KN83,NK90}.
Their model focused on processes of small 
cracks fusions into successively larger ones, 
accommodating the influence of mainshocks and aftershocks, 
juvenile crack genesis from tectonic stresses, 
crack healing, and anelastic-creep induced time 
delays, plus other effects.
Turcotte et al. \cite{Tur+99} have reinstated
this line of research considering a log-binned description 
of hierarchical aggregation and performing numerical tests 
to study its scaling properties.
Gabrielov et al. \cite{GNT99} first have
employed the Horton-Strahler hierarchical ranking
\cite{Horton45,NTG97}
to construct an exactly solvable model of a general 
inverse cascade process.
The Horton-Strahler ranks (see Sect. \ref{HS})
that came from hydrology and have been not well known
in physical applications
happened to be more natural than cluster masses 
(sizes, areas) in describing the aggregation process.
Moreover, the ranks are shown essential for 
formulating the analytical models \cite{GNT99}.
Recent efforts deal with studying the 
aggregation dynamics and its various scalings via exactly 
solvable hierarchical models and extensive simulations
\cite{MNTG04}.

Below we focus on the evolution of the first
spanning cluster in the the classical site-percolation 
model, and decribe it as a consecutive hierarchical
fusion of smaller clusters into larger ones.
Noteworthy, we are interested not in a final solution 
of a percolation state, but in an evolutionary path
leading from the juvenile single-particle clusters
to a self-similar population of clusters of arbitrary 
large size (limited by the finiteness of the lattice), 
the percolation cluster included.  
Thus we depart from the steady-state assumption of
\cite{GNT99,MNTG04,Tur+99} as well as from the 
asymptotic focus on the percolation onset typical
for the classical percolation studies \cite{SA}.

Specifically, we follow \cite{GNT99} and represent 
each cluster by a time-oriented tree that reflects the 
history of cluster formation.
The model dynamics is then described in terms of 
the corresponding trees using the well-developed
theory of hierarchical scaling complexities 
\cite{BP97,NTG97}.
An important role is played by the 
the Horton-Strahler scheme that provides a natural 
ranking for the tree-based structures.
Another important element is the Tokunaga classification 
that defines a special subclass of trees with
self-similar branching.
A large number of hierarchies observed in nature are
shown to be Tokunaga trees \cite{NTG97};
this is also the case for the clusters in percolation 
model \cite{GNT99,MNTG04}.  
We use the Tokunaga constraint together with
classical results on percolation dynamics (in 
terms of cluster masses) to
derive time-dependent scaling laws for
rank distribution of clusters.
Importantly, we report a three-exponent scaling for
the dynamics of a population of clusters of a given rank,
in deviation from the two-exponent scaling well-known
for the population of a given mass 
\cite{SA,Margolina+84}.
We also analyze deviations from the pure scaling
and confirm our results by numerical simulations.

The inverse cascades and aggregation (coagulation)
processes are important for evolution of
many natural hazardous processes:
earthquakes, landslides, and forest fires are argued 
to follow the hierarchical aggregation dynamics 
\cite{TMGR02,MNTG04}.
A general review of the theory and models of 
kinetics of irreversible aggregation is given by
Leyvraz \cite{Leyvraz03}.
An alternative approach to analytical modeling,
based on ideas from \cite{GNT99}, but using equations 
that are consistent with the mass action law of chemical
kinetics, can be found in 
da Costa et al. \cite{CGW02}.

The paper is organized as follows. 
The percolation model is described in Sect.~\ref{model};
this section also introduces tree representation of
clusters and the Horton-Strahler ranking.
In Sect.~\ref{MRD} we derive the average mass of 
clusters of a given rank using the Tokunaga constraint 
on cluster branching.
This result will be actively used in consecutive 
sections.
Section~\ref{RD} is devoted to the time-dependent
rank distribution of clusters.
First (Sect.~\ref{rankdist}), we establish the 
exponential rank distribution at percolation using 
the result of Sect.~\ref{MRD}.
We then proceed with time-dependent rank distribution;
Sect.~\ref{dynrankdist} introduces the three-exponent
scaling for ranks and compares it to the well-known
Stauffer's two-exponent scaling for cluster masses. 
Scaling for ranks averaged over the entire evolution
of the percolation cluster is derived in Sect.~\ref{AS};
this result is motivated by the heuristic studies
that typically use averaged observations on a system.  
Time-dependent finite-size corrections to the established 
scalings are described in Sect.~\ref{corrections}.
Our study of rank distributions is concluded in
Sect.~\ref{MDGR} by describig the time-dependent behavior 
of the total mass of clusters of a given rank.
Sect.~\ref{CFS} analyzes fractal properties of clusters 
and reports sharp increase of cluster fractal dimension
in the vicinity of percolation.
Sect.~\ref{DC} uses simulations to establish a 
notable constraint 
on the dynamics of a rank populations.
The results are discussed in Sect.~\ref{discussion}.

\section{Model}
\label{model}
\subsection{Dynamics}
We consider the classical 2{\it D} site-percolation
model \cite{SA}.
The model dynamics starts with an empty $L\times L$
square lattice. 
At each step a particle is dropped into a randomly 
chosen unoccupied site;  
thus each site can be either occupied by one and only 
one particle or empty. 
Two sites are considered {\it neighbors} if they share
one side; each site on a square lattice has four 
neighbors.
Cluster is defined as a group of occupied neighbor 
sites \cite{SA}.
Time refers to the steps at which particles drop
onto the lattice.
Since we do not have annihilation of particles, time
is formally equivalent to the number of particles on 
the lattice.
It is convenient to normalize time by the lattice
size $L^2$ so it varies from $\rho=0$ at the start to
$\rho=1$ when all sites are occupied. 
During the system evolution, occupied sites start to 
aggregate and clusters begin to form.
Once a sufficient number of particles is accumulated,
a percolation cluster is formed connecting the opposite
sides of the lattice vertically and/or horizontally.

The density $\rho$ increases monotonically from zero to 
its critical value $\rho_c$ at percolation. 
For an infinite lattice $\rho_c \approx 0.59274606$ \cite{NZ01}, 
while for a finite lattice it is smaller \cite{SA}:
\begin{equation}
\label{rhofin}
\rho_c(L) = \rho_c - cL^{-3}.
\end{equation}  

Many phenomena encountered in the percolation model 
mimic what we see when the phase transitions of the second 
kind occur.
Note however that these phenomena are of purely 
geometrical and statistical rather than physical 
nature.
Indeed, the physical percolation theory is largerly
predicated in this geometrical model and there are
many empirical links between them; this is why
the percolation model is said to be an example
of the {\it geometrical} phase transition of the second 
kind, and why its nomenclature emerges from that
of the physical critical phenomena.     

The theoretical description of the percolation 
dynamics is conventionally given in terms of 
the cluster masses \cite{SA};
and most of the universal scalings -- a benchmark
of phase transitions of second kind -- deal with 
parameters expressed via the mass distribution of 
clusters. 
However, if one is interested in analytical 
description of the aggregation process, the mass 
description happens to be inferior to the hierarchical 
rank approach \cite{GNT99,MNTG04}.
Properly defined ranks not only allow one to 
construct exactly solvable models of aggregation,
but also they are more feasible for observations in 
practice.
In addition, they reflect the individual history of 
cluster formation.
Below we follow the hierarchical approach 
of Gabrielov et al. \cite{GNT99} to study the 
percolation dynamics. 

\subsection{Tree representation of clusters}
\begin{figure}[p]
\centering\includegraphics{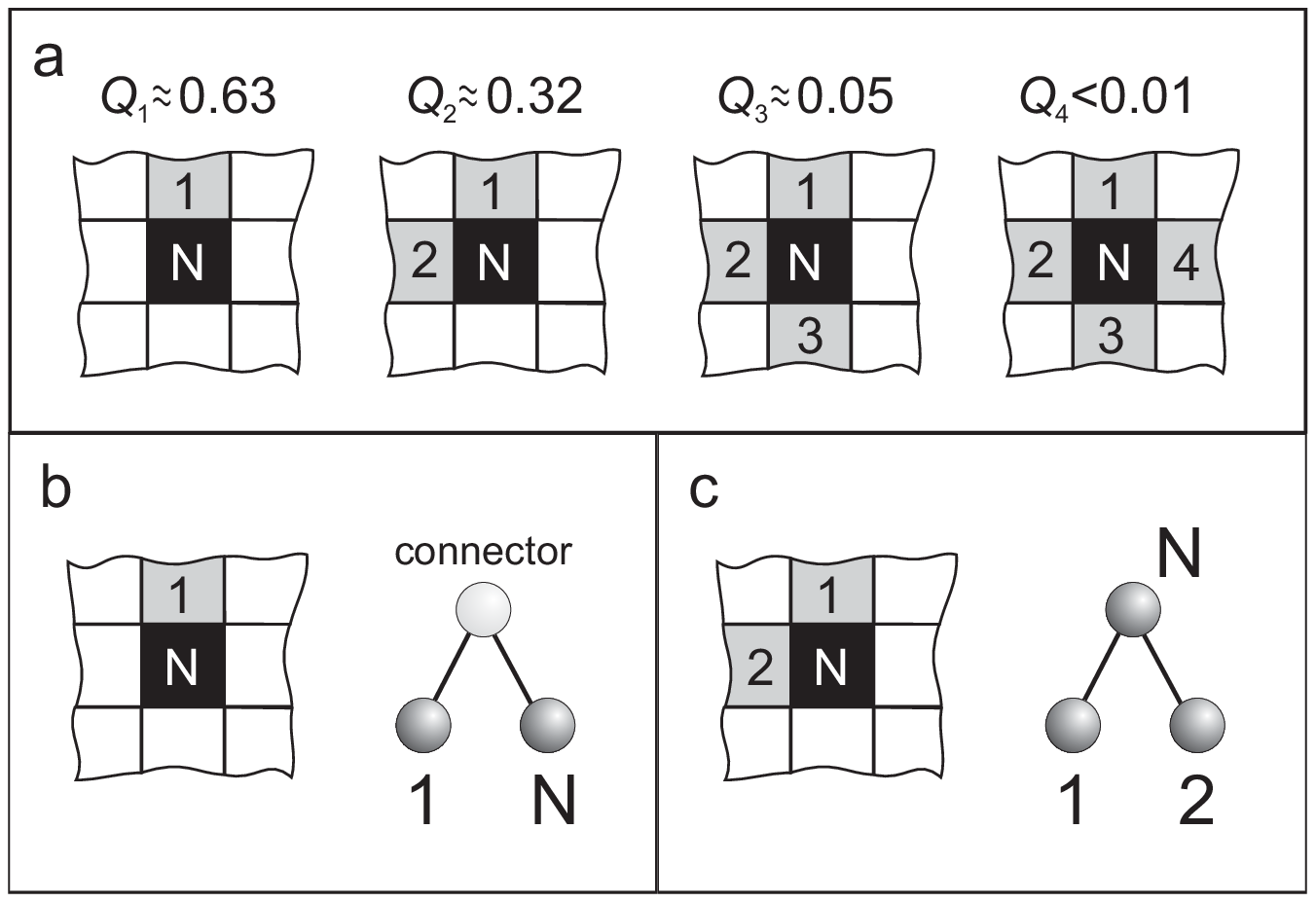}
\caption{Multiple coalescence of clusters.
a) Coalescence of clusters is materialized
by adding to the lattice a new particle 
{\bf N} (black) that is a neighbor to one, 
two, three, or four existing clusters 
(numbered gray sites).
The relative frequencies 
$Q_k$ , $k=1,2,3,4$
of $k$-coalescences
based on similations with 
$L=2,000$ are shown in the figure.
The corresponding tree is constructed as
shown in panel b) (for $k=1$) and c) (for $k=2$).
The cases $k=3,4$ are analogous to $k=2$.
Note that about 95\% of coalescences result
in merging two clusters. See text for details.}
\label{fig_model1}
\end{figure}

Each cluster in our model is represented by a 
tree that reflects the time-dependent formation 
of a cluster (its history), and is a subject for 
quantitative analysis.
Specifically, each one-particle cluster is 
represented by a trivial tree consisting of a 
single node.
When two clusters are merged together
their trees are also merged by adding a new
node (parent) for which they become children
(and siblings to each other.)
In our model, the coalescence of two or more clusters 
can only be materialized by adding to the lattice 
a new particle which will be a neighbor to one or
more existing clusters.
Figure~\ref{fig_model1}a illustrates the four 
possible types of coalescence.
We call $k$-coalescence in a situation when a newly
dropped particle (marked {\bf N} in the figure) is 
a neighbor to $k$ existing clusters (gray numbered
sites).
Numerical simulations on a square lattice with 
$L=2,000$ suggest the following relative frequencies 
$Q_k$ of $k$-coalescences: 
$Q_1\approx0.628$, $Q_2\approx0.318$, 
$Q_3\approx0.052$, $Q_4\approx0.002$.
Figures~\ref{fig_model1}b,c illustrate how
a tree is formed for different coalescence
types.
There are two basic situations: 
When a new particle is a neighbor to only
one existing cluster, it is considered as
an individual one-particle cluster that
is connected to the existing one.
The connecting node of the tree in this 
case does not
correspond to a particle on the lattice
(panel b).
When a new particle is droped in a neighbor to
two, three, or four existing clusters,
it is not condidered as an individual cluster.
Instead, it corresponds to the connecting
node in the tree (panel c).
Thus, the connecting node in a tree may or 
may not correspond to a lattice particle 
depending on the coalescence type. 
The branching parameter (number of children
for a given parent) of a tree for any 
cluster varies between 2 and 4.  
Note that both 1- and 2-coalescences result
in merging only two clusters; 
accordingly, most of the observed coalescences 
(about 95\%) involve only two clusters while 
coalescence of three or four clusters is 
extremely rare.

\begin{figure}[p]
\centering\includegraphics{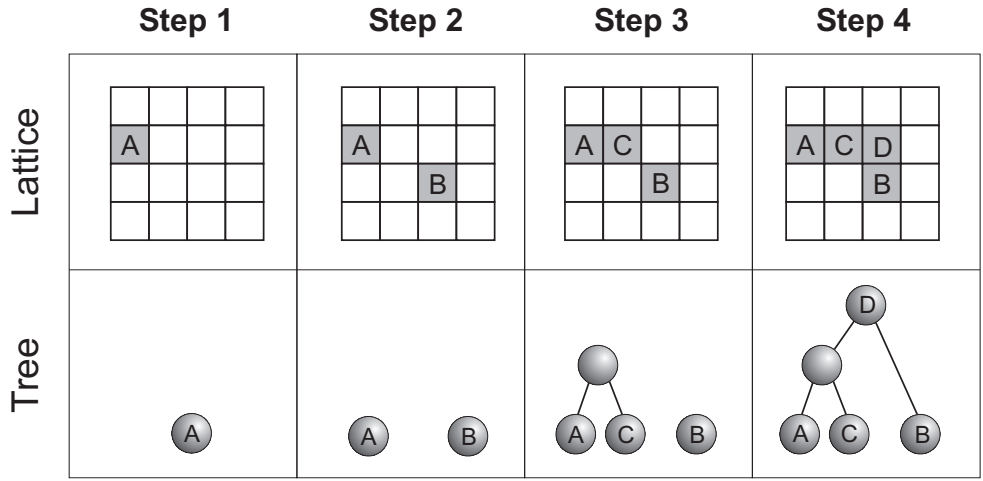}
\caption{Tree representation of clusters: scheme. 
The dynamics is from left to right. 
At first step particle {\bf A} is dropped onto 
the lattice and a one-particle cluster
is formed; it is represented by a 
one-node tree.
At second step another one-particle 
cluster {\bf B} is formed; it is represented
by another one-node tree. 
At third step new particle {\bf C} coalesces with 
cluster {\bf A} to form two-particle cluster 
{\bf AC}. 
This cluster is represented by a three-node tree;
note that the connecting node of the tree does not 
correspond to any particle.
At fourth step new particle {\bf D} connects 
existing clusters {\bf AC}
and {\bf B} to form four-particle cluster {\bf ABCD}.
This cluster correspond to a five-node tree.}
\label{fig_model2}
\end{figure}

\begin{figure}[p]
\centering\includegraphics{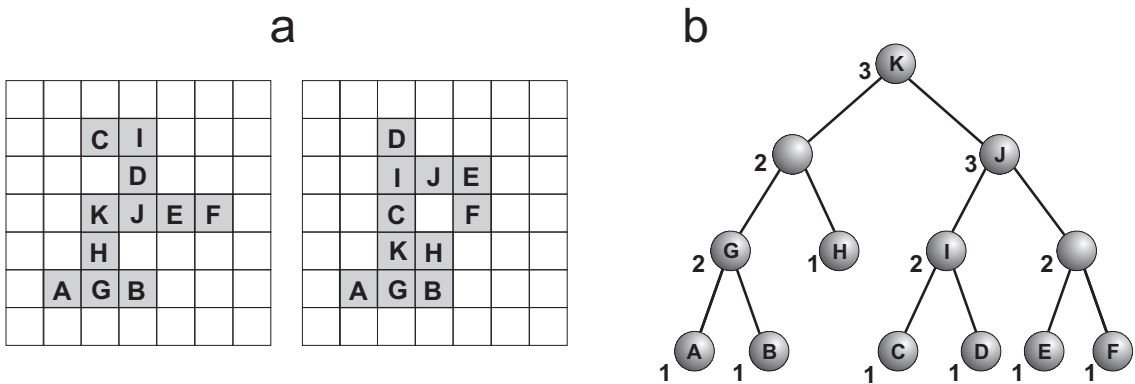}
\caption{a) Non-uniqueness of tree representation. 
Two different 11-particle clusters that 
correspond to the same tree shown in panel b). 
Particles have been dropped according to their
alphabet marks; so first was the particle {\bf A},
then {\bf B}, etc.
b) Horton-Strahler ranking: illustration.
The ranks are shown next to the tree nodes.}
\label{fig_model3}
\end{figure}

The consecutive process of tree formation
for a simple four-particle cluster
is illustrated in Fig.~\ref{fig_model2}.
Importantly, the individual evolution of a 
cluster is crucial in constructing the corresponding 
hierarchical tree. 
To construct the tree one needs to consider 
all consecutive coalescences that have formed 
the cluster, not only its final shape.
Therefore, it is clear that the same tree may correspond to 
clusters of different shape: Figure~\ref{fig_model3}a
shows two 11-particle clusters that both
correspond to the same tree shown in panel b.
Therefore, working with trees, we unavoidably
narrow the information about the cluster
population.
Notice however both trees capture an excessively 
larger amount of information than mere cluster 
masses.
Summing up, the time evolution of a cluster
is neccesary and sufficient to uniquely
determine the corresponding tree, while
the inverse is not true.
The problems of describing the set of trees
that might correspond to a given cluster,
and the set of clusters that correspond to
a given tree is beyond the scope of this
paper.
Next, we describe the ranking of clusters,
presenting a conventional alternative to
the logarithmic binning of cluster masses.

\subsection{Horton-Strahler ranking}
\label{HS}
The appropriate ordering of trees (clusters) 
is very important for meaningful description and 
analysis of the model dynamics. 
The problem of such an ordering is not trivial
since the clusters may grow and coalesce
in a variety of peculiar ways.
An advantageous way to solve this problem is
given by the Horton-Strahler topological 
classification of ramified patterns
\cite{Horton45,Strahler,BP97} 
illustrated in Fig.~\ref{fig_model3}b:
One assigns ranks to the nodes of a tree, 
starting from $r=1$ at leaves (clusters 
consisting of one particle.)
When two or more clusters with ranks $r_i$,
$i=1,\dots,n$ 
merge together, a new cluster is formed
with the rank \cite{BP97}:
\[r=\left\{\begin{array}{ll}
r_1+1,&\mbox{ if }r_i=r_1~\forall~i=1,\dots,n\\
\max{(r_i)},&\mbox{ otherwise.}
\end{array}\right.
\]
The rank of a cluster is that of the root
of the corresponding tree.
It is possible to consider an alternative 
definition of ranks: 
When at least two clusters with
rank $r$ coalesce, and other coalescing 
clusters have a lower rank, the rank of a new 
cluster becomes $r+1$.
Clearly, the two definition coincide when
only two clusters coalesce.
The results reported in this paper are
independent of the particular definition,
since coalescence of more than two clusters 
(especially of high ranks) is a rare event.

Originally introduced in geomorphology 
by Horton \cite{Horton45} and later refined by 
Strahler \cite{Strahler}, this classification 
is shown to be inherent in various geophysical, 
biological, and computational applications 
\cite{BP97,GNT99,MNTG04,NTG97,Toro01,TPN98}.

\section{Mass-rank distribution}
\label{MRD}
In this section we derive the distribution of the 
average mass $m_r$ of rank $r$ clusters.
It will be used consequently 
to connect various mass and rank scaling laws.
First, we define the branching ratio $T_{ij}$ 
for a given cluster (tree) as the number 
$N_{ij}$ of subclusters (nodes) of rank $i$ that 
joined subcluster (node) of rank $j$, averaged
over subclusters (nodes) of rank $j$ 
\cite{NTG97,Tokunaga}:
\[T_{ij}=\frac{N_{ij}}{N_j}.\]

Next we note that the mass of a rank $r$ cluster 
is the sum of two $r-1$ cluster masses that
formed the cluster (we ignore the possibility for
three or more clusters to coalesce at the same step), 
plus a unit mass of a joining particle, plus the 
mass of all the lower-rank clusters that joined the
considered cluster, hence:  
\begin{eqnarray}
\label{mr}
m_1&=&1\nonumber\\
m_2&=&(2\,m_1+P)+T_{12}(m_1+P)\nonumber\\
m_3&=&(2\,m_2+1)+T_{23}(m_2+1)+T_{13}(m_1+P)\nonumber\\
&\dots&\nonumber\\
m_k&=&(2\,m_{k-1}+1)+\sum_{i=1}^{k-1}T_{k-i\,k}(m_{k-i}+1)
-(1-P)T_{1\,k},~k\ge 3.
\end{eqnarray}
Here the coefficient $P$ addresses the
possibility for a one-particle cluster to join 
another cluster in two ways: via a 
one-particle connector (with probability $P$)
or directly (with probability $1-P$); the
clusters with $r>2$ can only join other clusters
using a one-particle connector.

It was predicted by Gabrielov et al.
\cite{GNT99} and later confirmed by 
simulations \cite{MNTG04} that clusters
in percolation model obey the Tokunaga
scaling \cite{Tokunaga} asymptotically in $k$:
\begin{equation}
\label{tokunaga}
T_{i\,i+k}=T_k=s_0s^{k-1}.
\end{equation}
This rewrites Eq.~(\ref{mr}) for $k\ge 3$ as
\[m_k=(2\,m_{k-1}+1)+\sum_{i=1}^{k-1}T_{i}(m_{k-i}+1)
-(1-P)T_{k-1}.\]
Assuming the mass-rank relation in the form
$m_r=c^{r-1}$, $c>1$ we obtain
\begin{eqnarray}
\label{mr1}
c\,^{k-1}&=&2c\,^{k-1}+1+\sum_{i=1}^{k-1}s_0s\,^{i-1}
\left(c\,^{k-i-1}+1\right)-
(1-P)s_0\,s\,^{k-1}\nonumber\\
&=&c\,^{k-2}\left[
2+\frac{1}{c\,^{k-2}}
+s_0\frac{1-\left(s/c\right)^{k-1}}
{1-s/c}
+\frac{s_0}{c\,^{k-2}}\,\frac{s\,^{k-1}-1}{s-1}
-(1-P)s_0\left(s/c\right)^{k-2}\right].\nonumber
\end{eqnarray}
It is easily checked that this equation 
has a solution only if $c>s$; 
thus $s/c<1$ and for large $k$ then follows
\[c^{k-1} = c^{k-2}\left[2+\frac{s_0}{1-s/c}\right]\] 
leading to the final equation
\[c^2-c(2+s+s_0)+2s=0\]
with solution:
\begin{equation}
\label{c}
c=\frac{2+s+s_0\pm\sqrt{(2+s+s_0)^2-8s}}{2}.
\end{equation}

Remarkably, the model of Gabrielov et al. \cite{GNT99}
predicts in a Euclidean (assuming clusters of regular,
non-fractal, shape) limit of an inverse cascade model
\[s_0\approx 0.55495813,~~ s=1/s_0\approx1.80193774,
~~{\rm and~~} c=1/s_0^2\approx3.24697602.\]
The Eq.~(\ref{c}) in this case gives 
$c(s_0,s)=3.24697960$ (this is the only
solution such that $c>s$), 
which is remarkably close (6 digits) to 
the result of \cite{GNT99}.  
Furthermore, the non-Eucledian (assuming fractal 
shape of clusters) steady-state simulations 
of Morein et al. \cite{MNTG04} suggest
\[s\approx3.0253, \qquad s_0\approx0.6993,
\qquad c\approx4.325,\]
which exactly solves Eq.~(\ref{c}).
We found it quite amazing that our complimentary 
set of assumptions used to derive (\ref{c}) lead 
to the same numerical results as analytical study 
\cite{GNT99} and simulations of \cite{MNTG04}.
This suggests an underlying connection between
our approaches to describe the hierarchical 
aggregation.

\begin{figure}[p]
\centering\includegraphics{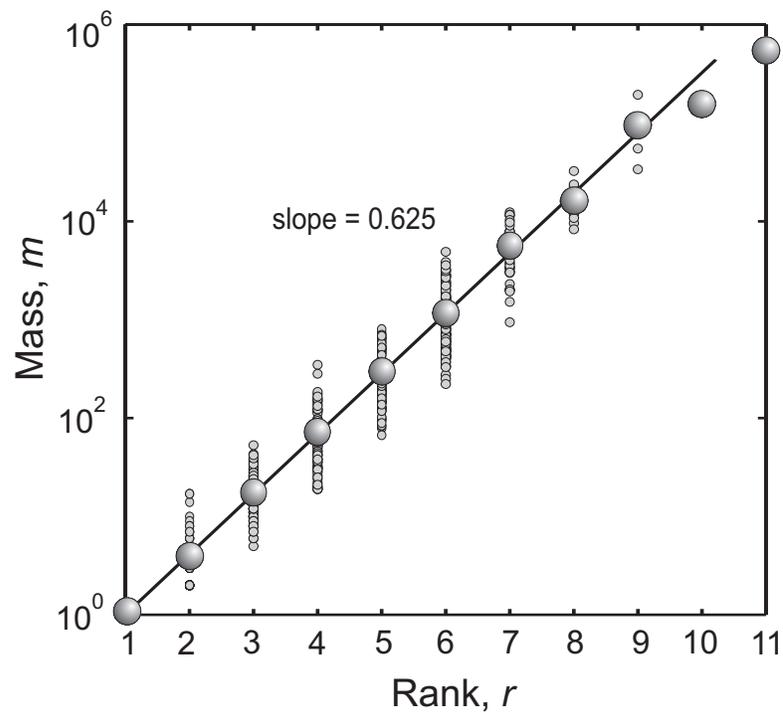}
\caption{Mass-rank distribution observed
on a 2,000$\times$2,000 lattice at percolation.
Dots -- individual clusters, balls -- average
mass $m_r$ within a given rank.
Line shows the relation 
$m_r=\left[10^{\,0.625}\right]^{\,r-1}=4.2^{\,r-1}$.}
\label{fig_MR}
\end{figure}

The observed mass-rank distribution of clusters 
at percolation is shown in Fig.~\ref{fig_MR}; 
it obeys the exponential relation
\begin{equation}
\label{MR}
m_r= 10^{\,\gamma (r-1)} = c^{\,r-1},
\end{equation}
with 
$\gamma\approx 0.625$,
$c=10^{\gamma}\approx 4.2$.
Our simulation suggest that the mass distribution 
within a given rank is approximately lognormal 
(not shown) with the mean given by Eq.~(\ref{MR}) and 
a rank-independent standard deviation. 

The relation (\ref{MR}) is a key
element in our further analysis. 
As we will show, the distribution of cluster 
ranks at percolation (Sect.~\ref{rankdist}) 
and its finite-size corrections 
(Sect.~\ref{corrections})
are obtained from the corresponding 
classical laws for masses by simple substituting 
the relation (\ref{MR}).
At the same time, one of the most important
results: the time dependent rank distribution can
not be obtained this way and requires an 
additional treatment (Sect.~\ref{dynrankdist}).  

\begin{figure}[p]
\centering\includegraphics{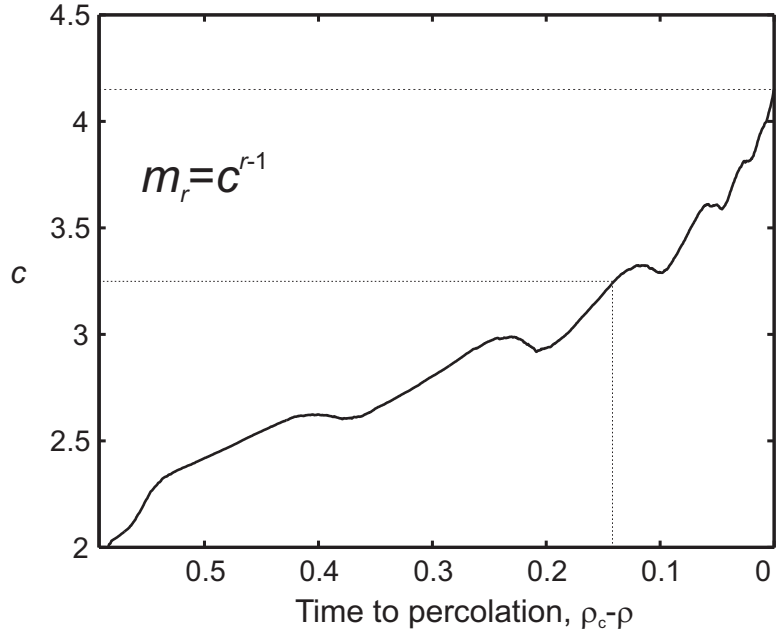}
\caption{Parameter $c$ of the mass-rank relation
$m_r=c^{r-1}$ as a function of time.
At percolation $c(\rho_c)\approx 4.2$;
the Euclidean limit of \cite{GNT99} corresponds
to $c=3.25$, it is reached at 
$\rho_c-\rho\approx 0.14$.}
\label{fig_c}
\end{figure}

The exponential relation of Eq.~(\ref{MR}) happens 
to be valid over the entire time interval 
$0<\rho\le\rho_c$.
The corresponding dynamics of $c(\rho)$ is shown in
Fig.~\ref{fig_c}: 
it grows with time from about 2.0 at the earliest 
stages to 4.2 at percolation.
This growth reflects the fact that clusters 
become more weighty with time due to coupling
with the clusters of lower ranks (which does not
change the rank but increases the mass).  
The growth is not monotonous; it is accompanied
by pronounced log-periodic oscillations which
are associated with creation of new ranks.
The log-periodic oscillations that accompany
general power-law increase of observed parameters
have been found in many systems including 
hierarchical models of defect development
\cite{NTG95},
biased diffusion on random lattices \cite{SS98},
diffusion-limited aggregation (DLA) \cite{Sor+96},
and others.
Log-periodic oscillations 
can be naturally explained by the
Discrete Scale Invariance (DSI) \cite{Sornette04},
which occurs in a system whose observables 
scale only for a discrete set of values.
A famous example of DSI is given by the Cantor set 
that pocesses a discrete scale symmetry:  
In order to superimpose its scaled image onto the 
original, one has to stretch it by the discrete factors 
$3^n$, $n=1,2,\dots$, not a continuous set of values.  
The Cantor set and percolation belong to systems with 
built-in geometrical hierarchy, leading to DSI.  
In our particular system, ranks take only a 
countable set of values.  
Creation of new ranks necessarily disrupt the 
system in a discontinuous way resulting in 
the log-periodicity. 

Now we return to the numerical value of 
parameter $c$. 
In steady-state simulations of \cite{MNTG04}
$c = 4.325$, which is reasonably close to what we
observe at percolation.
Recall that the models of \cite{GNT99,MNTG04} use
the ``fractal correction'' $\epsilon$ to the cluster 
shape; this correction affects the rate $r_{ij}$ of
clusters coalescence:
\[r_{ij}\approx \epsilon^{-|j-i|}L_iL_j,\]
where $L_i$ is the total boundary size of 
the clusters of rank $i$.
The correction $\epsilon$ can be expressed as
\[\epsilon=\frac{1}{\sqrt{c}}\,\frac{c-1}{c-2},\]
which, together with results of Fig.~\ref{fig_c}, 
shows that in the percolation model
$\epsilon$ decreases in time passing the 
Euclidean limit $\epsilon=1$ \cite{GNT99} at 
$(\rho_c-\rho)\approx 0.14$ and approaching 
the steady-state ``fractal'' 
$\epsilon\approx 0.68$ \cite{MNTG04} at
$\rho=\rho_c$.
The interval $2< c \le 4.2$ observed
during $0<\rho\le \rho_c$ corresponds 
to $0.68\le\epsilon<\infty$. 

\section{Rank distribution}
\label{RD}
This section is devoted to establishing
various time-dependent scaling laws for clusters
of a given rank.
We will see that it is tipically impossible 
to derive such laws by applying the mass-rank 
relation (\ref{MR}) to the coresponding
well-known laws for cluster masses.
This illustrates an original character and 
richness of the rank description and prompts for 
developing new methods of analysis.
We start with the simplest problem: 
rank distribution at percolation.

\subsection{Distribution at percolation}
\label{rankdist}
We start recalling the well-known cluster mass 
distribution at percolation \cite{SA}:
\begin{equation}
\label{GRm}
n_m(\rho_c)\sim q_0\,m^{-\tau},
\end{equation}
where $n_m(\rho_c)$ is the number of clusters of
mass $m$ per lattice site, 
and the Fisher exponent $\tau=187/91\approx 2.05$ 
is universal for 2{\it D} systems \cite{Fisher67,SA}.
Figure~\ref{fig_GRm} illustrates the mass
distribution at percolation for a system with $L=2000$;
to smooth out statistical fluctuations it shows the 
number of clusters with mass equal to or larger than $m$:
$\sum_{m'\ge m} n_m(\rho_c)$.
Equation~(\ref{GRm}) suggests the slope
$\tau-1\approx\,1.05$, while the observed slope
$0.96$ is somewhat less than that.
This is due to the impact of two concurrent
phenomena: so-called ``deviation from scaling''
at small $m$ \cite{Hoshen+79} and finite-size effects
at large $m$ \cite{Margolina+84,Hoshen+79};
they are discussed below in Sect.~\ref{corrections}.

Now, we use Eq.~(\ref{GRm}) to derive the distribution of 
the number $n_r(\rho_c)$ of the clusters of rank $r$ at 
percolation.
Taking summation over all clusters of rank $r$ and mass $m$
we obtain:
\begin{eqnarray}
\label{nrm0}
n_r(\rho_c)&=&\sum n_{r,m}(\rho_c)
=q_0\,\sum_{m_{\rm lo}}^{m_{\rm up}}\,m^{-\tau}\nonumber\\
&\sim&\frac{q_0}{\tau-1}
\left[(m_{\rm lo})^{-\tau+1}-(m_{\rm up})^{-\tau+1}\right]\nonumber\\
&=&\frac{q_0}{\tau-1}
\left[\left(\frac{m_{\rm lo}}{  m_r  }\right)^{-\tau+1}
-\left(\frac{m_{\rm up}}{  m_r  }\right)^{-\tau+1}\right]
m_r^{-\tau+1}.
\end{eqnarray}
Our simulations suggest (not shown) that the mass distribution 
within a given rank is lognormal with a rank-independent 
standard deviation.
Thus, for arbitrary upper and lower quantiles 
$m_{\rm up}$, $m_{\rm lo}$ of this distribution
the values 
\[\frac{m_{\rm lo(up)}}{m_r}\]
are rank independent.
Using this, we finally express $n_r(\rho_c)$ via $m_r$:
\begin{equation}
\label{nrm}
n_r(\rho_c)=p_0\,  m_r  ^{-\tau+1}
\propto   m_r  ^{-1.05}.
\end{equation}

The power law (\ref{nrm}) is observed in a steady-state
aggregation model of \cite{MNTG04} with index $1.147$.
This index increase comparing to our $1.05$ is due to 
the fact that in \cite{MNTG04}
intermediate clusters are removed from the lattice
providing extra space for a larger number of smaller 
clusters. 

Combining the mass-rank relation (\ref{MR}) with
(\ref{nrm}) we obtain the following 
exponential rank distribution at percolation:
\begin{eqnarray}
\label{GRr}
n_r(\rho_c)&\sim& p_0\,  m_r  ^{-\tau+1}
=p_0\,\left(c^{-\tau+1}\right)^{r-1}
= p_1\,10^{-b\,r}
\end{eqnarray}
with
\[p_1=p_0\,c^{\tau-1},~b=(\tau-1)\,\log_{10}c\approx 0.62.\]

This is indeed what we observe in Fig.~\ref{fig_GRr}
where the rank distribution $n_r$ at percolation
is shown by the dash-dotted line.
The study \cite{MNTG04} suggests $c^{1-\tau}=0.186$
while our predictions and observations lead to 
$c^{1-\tau}\approx 4.2^{-1.05} = 0.22$.
The two values are in good agreement, the slight 
difference is explained, as in 
Eq.~(\ref{nrm}), by removal of intermediate
clusters in \cite{MNTG04}.
Next we consider the rank distribution for $\rho\ne\rho_c$.

\subsection{Dynamical rank distribution: three-exponent scaling}
\label{dynrankdist}
Here we expand results of the previous section
by establishing the time-dependent rank dustribution.
First, we consider the dynamics of rank population. 

\subsubsection{Temporal dynamics of rank population}
The dynamics of the total number $(n_r\cdot L^2)$ of the 
clusters of a given rank $r$ is illustrated in 
Fig.~\ref{fig_rank} for $r=5,6,7$.
The population follows a characteristic bell-shaped 
trajectory, with percolation at its rightward limb. 
As in the case of mass description, one does not observe 
steady-state behavior in the cluster dynamics:
The population of each rank steadily develops to its peak
as a result of merging of the clusters of lower ranks; 
then it starts decreasing, giving birth to the 
clusters of higher ranks. 
As naturally follows from the model definition, the
peak of the population of a higher rank comes 
after the peak of a lower rank.
Figure~\ref{fig_rank_all} shows the population dynamics
for the ranks $1\le r \le 11$ in semilogarithmic scale.
Here one clearly sees the similarity in the dynamics
of different ranks.
Note that this figure is remarkably similar to
Fig.~7 from \cite{Tur+99} that shows the dynamics of
clusters with logarithmically binned masses. 
We now proceed by establishing the appropriate 
time-dependent scaling laws.

\subsubsection{Time-dependent mass distribution}
Recall that the temporal dynamics of the cluster mass 
distribution is given by the two-exponent scaling law 
\cite{Stauffer75,SA,Margolina+84}:
\begin{equation}
\label{2exp}
n_m(\rho)\sim m^{-\tau}\,f_0(z),\qquad 
z=(\rho_c-\rho)m^{\sigma}+z_0,
\end{equation}
with $\sigma=1/2$. 
The function $f_0$ has a bell-shaped form with maximum
to the left of percolation;
it can be roughly approximated by a Gaussian function
\cite{Hoshen+79,Margolina+84}:
\begin{equation}
\label{f0}
f_0(z)\propto \exp\left(-a\,z^2\right).
\end{equation}
Note that the shift $z_0$ is independent 
of $m$.

Considered as a function of $m$, the two-exponent 
scaling explains the power law mass distribution 
(\ref{GRm}) at percolation (with $q_0=f_0(z_0)$) 
as well as the downward bend for $\rho<\rho_c$, 
clearly observed in Fig.~\ref{fig_GRm} (dashed line);
while as a function of $\rho$ it describes the 
bell-shaped dynamics of clusters with given mass $m$.

\subsubsection{Time-dependent rank distribution}
\label{rscale}
Combining the scaling laws (\ref{MR}) and (\ref{2exp})
one formally obtains the two-exponent scaling for 
rank dynamics.
However, the two exponent scaling does not 
work for ranks; to show this we assume more generally
\begin{equation}
\label{r2exp}
n_r(\rho)\sim g_0(z)10^{-br},~
z=(\rho_c-\rho)h(r)+z'_0,
\end{equation}
which is consistent with the exponential rank distribution
of Eq.(\ref{GRr}) at percolation (with $p_0=g_0(z'_0)$). 
Possible deviations from the pure exponential law at 
$\rho<\rho_c$ (clearly observed in Fig.~\ref{fig_GRr}) 
and dynamics of a given rank 
(see Figs.~\ref{fig_rank},\ref{fig_rank_all}) 
are described by specific form of the functions 
$g_0(\cdot)$ and $h(\cdot)$.
Following \cite{Hoshen+79} we define  
\begin{equation}
\label{nu}
\nu_r(z):=\frac{n_r(z)}{n_r(z'_0)}
=\frac{g_0(z)}{g_0(z'_0)}.
\end{equation}
and choose $h(\cdot)$ in such a way that positions of 
the peaks of $\nu_r(z)$ coincide for different $r$;
it is always possible by choosing the appropriate 
time change $h(r)$.
Figure~\ref{fig_nu}a shows the ratio $\nu_r(z)/\nu_1(z)$
for $r=2,3,6,8$.   
One can see that the two-exponent scaling
does not work in our case: the curves do not coincide.

Nevertheless, the simple scaling picture is restored 
by introducing the additional, third, shift exponent:
\begin{equation}
\label{r3exp}
h(r)= a_1\,10^{\sigma_1\,r},\qquad 
z'_0(r)=a_2\, 10^{-\sigma_2\,r}.
\end{equation}
Function $g_0$ still can be approximated by a 
Gaussian function
\begin{equation}
\label{g0}
g_0(z)\propto \exp\left(-\frac{z^2}{2}\right).
\end{equation}

Once the correct scaling form is established, 
the use of (\ref{MR}) is again legitimate, and
the exponent $\sigma_1$ in Eq.~(\ref{r3exp}) 
can be evaluated as:
\[\sigma_1=\sigma\log_{10} \hat{c}
\approx 0.24,\]
where $\hat{c}\approx 3$ is the median of 
$c$ values observed during $\rho<\rho_c$. 
The observed exponent $\sigma_1\approx 0.23$ 
(not shown) is fairly close to its predicted 
value.
The shift exponent is estimated as 
$\sigma_2\approx 0.03$; while 
scale coefficients are $a_1\approx 1.54$,
$a_2\approx 1.43$. 
The function $g_0(z)$ that uses these estimates 
is shown in Fig.~\ref{fig_g0} where different symbols 
depict clusters of different ranks.
The collapse is obvious, confirming the validity
of the three-exponent scaling (\ref{r2exp}), (\ref{r3exp}),
(\ref{g0}).

In the scaling for cluster masses, the time renormalization 
$(\rho_c-\rho)m^{\sigma}$ collapses the dynamics 
of mass $m$ clusters onto the master 
curve $f_0(z-z_0)$ with its only peak shifted by 
$z_0$ leftward from percolation; 
the shift $z_0$ is mass independent.
Similarly, in the scaling for ranks the time renormalization
$(\rho_c-\rho)10^{\sigma_1\,r}$ collapses the dynamics
of rank $r$ clusters onto the master curve $g_0(z-z'_0)$,
although the shift now is rank dependent and is given
by $10^{\sigma_2\,r}$.
To illustrate this, we show the position of percolation 
on the righthand limb of the Gaussian $g_0(z-0.51)$ 
in Fig.~\ref{fig_nu}b.
The higher the rank, the closer the position of percolation
to the peak of $g_0$.

\subsection{Averaged scaling}
\label{AS}
In applications, it is often impossible to
measure the size distribution of system 
elements at a given time instant.
Moreover, sometimes the instantaneous size 
distribution does not exist at all: This is 
indeed the case for the systems described by 
marked point processes widely used to model 
seismicity, volcano activity, starquakes, etc. 
\cite{DVJ}.
In such situations one uses the averaged 
measurements.
For instance, the famed Gutenberg-Richter law 
\cite{GR54,Tur97,BZ03} that gives exponential
approximation to the size distribution of 
earthquakes (via their magnitudes)
is valid only after appropriate averaging 
over a wide spatio-temporal domain.
This explains the importance of the question:
How do the distributions of Eq.~(\ref{2exp}),
(\ref{r2exp}) change after temporal averaging?

We answer this question for averaging
over $0\le\rho\le\rho_c$.
For the mass distribution this leads to:
\begin{eqnarray}
\label{avem}
\widehat{n_m}&:=&\int_0^{\rho_c} n_m(\rho)d\rho
=\int_0^{\rho_c} f_0(z)\,m^{-\tau}d\rho\nonumber\\
&\propto&\int_0^{\rho_c}\exp\left\{-a\left[(\rho_c-\rho)m^{\sigma}
-z_0\right]^2\right\}m^{-\tau}d\rho\nonumber\\
&\propto& m^{-\tau-\sigma}\int_{u_1}^{u_2}
\exp\left\{u^2/2\right\}du\nonumber\\
&\propto& m^{-\tau-\sigma}~~(\approx m^{-5/2}).
\end{eqnarray}
Here the last step neglects the weak dependence
of the integral on $m$ (and uses the values 
$\tau\approx 2.0$, $\sigma=1/2$).
The validity of (\ref{avem}) is confirmed by the
observed averaged mass distribution shown by 
the solid line in Fig.~\ref{fig_GRm}.
The averaged mass distribution is similar
to that at percolation:
it retains the power-law form while the 
slope is increased by $1/2$ due to averaging.

Similarly, we obtain for ranks:
\begin{eqnarray}
\label{aver}
\widehat{n_r}&:=&\int_0^{\rho_c} n_r(\rho)d\rho
=\int_0^{\rho_c} g_0(z)\,10^{-br}d\rho\nonumber\\
&\propto&\int_0^{\rho_c}\exp\left\{-a'\left[
(\rho_c-\rho)10^{\sigma_1\,r}-a\,10^{-\sigma_2\,r}\right]^2
\right\}10^{-b\,r}d\rho\nonumber\\
&\propto& 10^{-(\sigma_1+b)\,r}\int_{u_1}^{u_2}
\exp\left\{u^2/2\right\}du\nonumber\\
&\propto& 10^{-(\sigma_1+b)\,r} = 
10^{r\,(1-\sigma-\tau)\log_{10}\tilde{c}}=
10^{-r\,\alpha_r}.
\end{eqnarray}
The exponent $\alpha_r$ may vary from 0.71 to 0.93 
depending on $3.0\le\tilde{c}\le 4.2$ (the range 
of $c$ values for the time when at least
three ranks have been formed so the estimation of 
the distribution slope is meaningful). 
Simulations suggest (solid line in Fig.~\ref{fig_GRr})
$\alpha_r=0.87$, which is in good agreement with
our prediction.
Again, the averaged rank distribution retains the 
exponential form of the distribution at percolation;
while its index has increased due to averaging.

\subsection{Correction to simple scaling}
\label{corrections}
Due to finiteness of the lattice,
the results of previous sections require 
some corrections to match exactly the simulated 
rank distributions.
The appropriate corrections are described below.  

\subsubsection{Corrected scaling at percolation}

\begin{figure}[p]
\centering\includegraphics{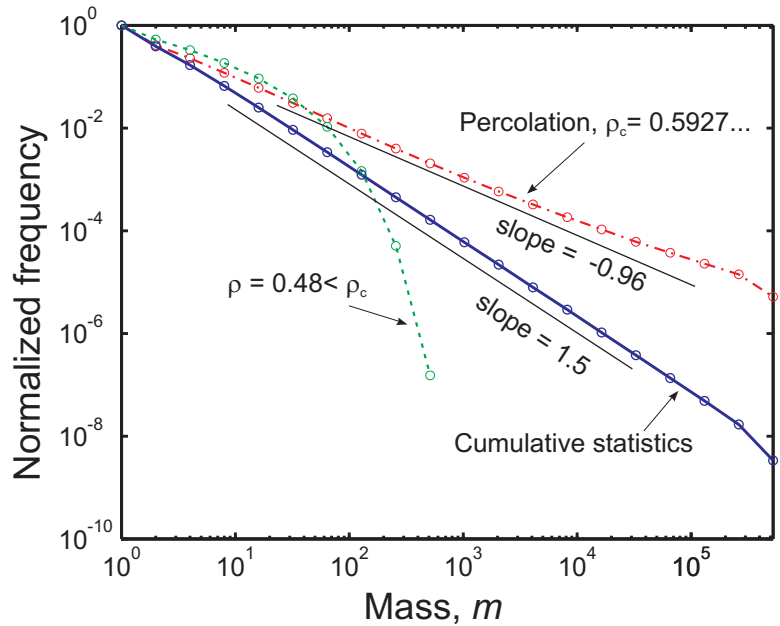}
\caption{Mass distribution of clusters
observed on a $2,000\times 2,000$ lattice
at percolation $\rho=\rho_c$ (dash-dotted line),
$\rho=0.48$ (dashed line), and averaged
over $0<\rho<\rho_c$ (solid line).
To smooth out statistical fluctuations
we show the cumulative distribution: 
$\propto\sum_{m'>m}n_m$.
For comparison, all curves are normalized 
to unity at $m=1$.}
\label{fig_GRm}
\end{figure}

\begin{figure}[p]
\centering\includegraphics{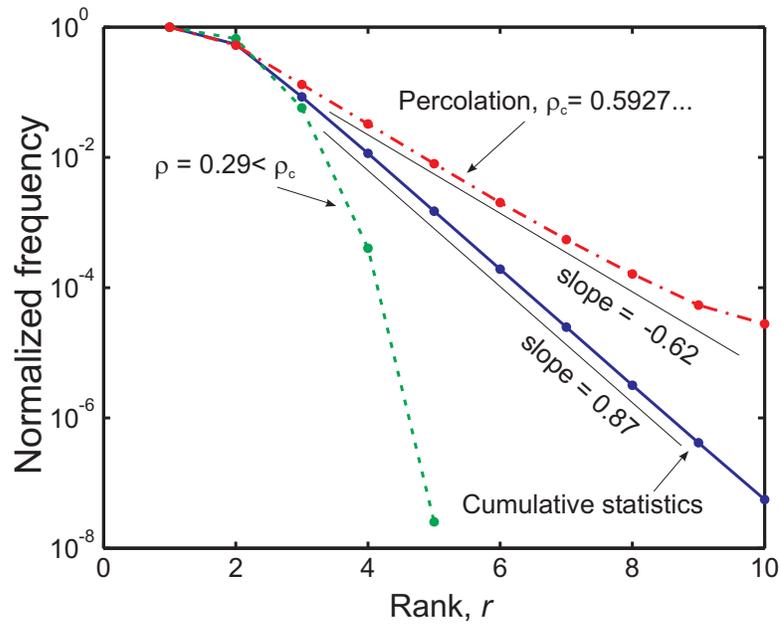}
\caption{Rank distribution of clusters
observed for $2,000\times 2,000$ lattice
at percolation $\rho=\rho_c$ (dash-dotted line),
$\rho=0.29$ (dashed line), and averaged
over the percolation cycle $0<\rho<\rho_c$ 
(solid line).
For comparison, all curves are normalized 
to unity at $r=1$.}
\label{fig_GRr}
\end{figure}

\begin{figure}[p]
\centering\includegraphics{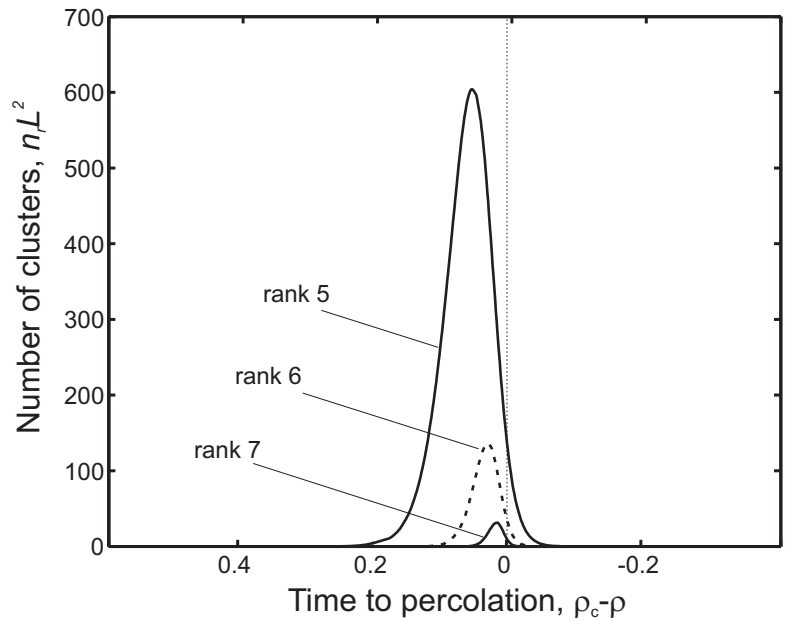}
\caption{Dynamics of population $n_r\cdot L^2$ of a 
given rank, $r=5,6,7$ for $L=2,000$.
Moment of percolation is depicted by
a vertical dashed line.}
\label{fig_rank}
\end{figure}

\begin{figure}[p]
\centering\includegraphics{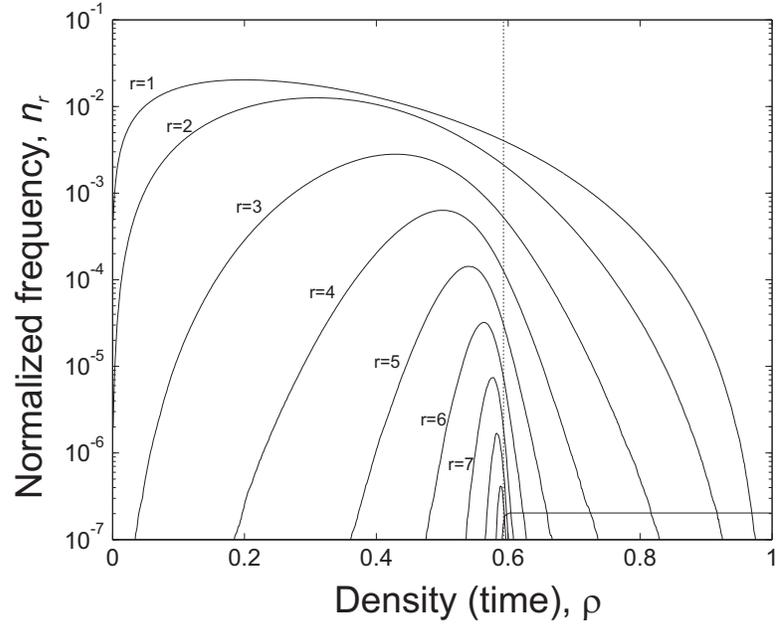}
\caption{Dynamics of population $n_r$ of a 
given rank, $1\le r\le11$ in semilogarithmic scale.
Moment of percolation is depicted by
a vertical dashed line. 
(Cf. Fig.~7 from \cite{Tur+99}).}
\label{fig_rank_all}
\end{figure}

\begin{figure}[p]
\centering\includegraphics{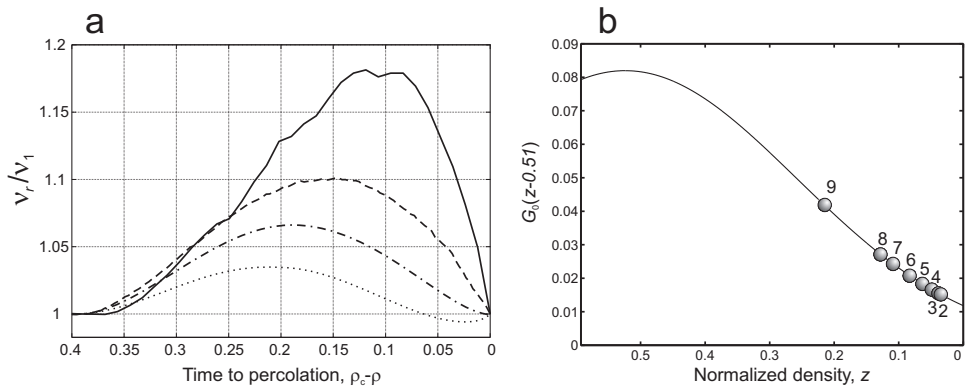}
\caption{Scaling for rank dynamics.
a) Ratios $\nu_r(z)/\nu_1(z)$ do not collapse
thus rejecting the two-exponent scaling
hypothesis; see details in Sect.~\ref{rscale}.
b) Position of percolation on the normalized
Gaussian $g_0(z-0.51)$; see details in 
Sect.~\ref{rscale}.}
\label{fig_nu}
\end{figure}

\begin{figure}[p]
\centering\includegraphics{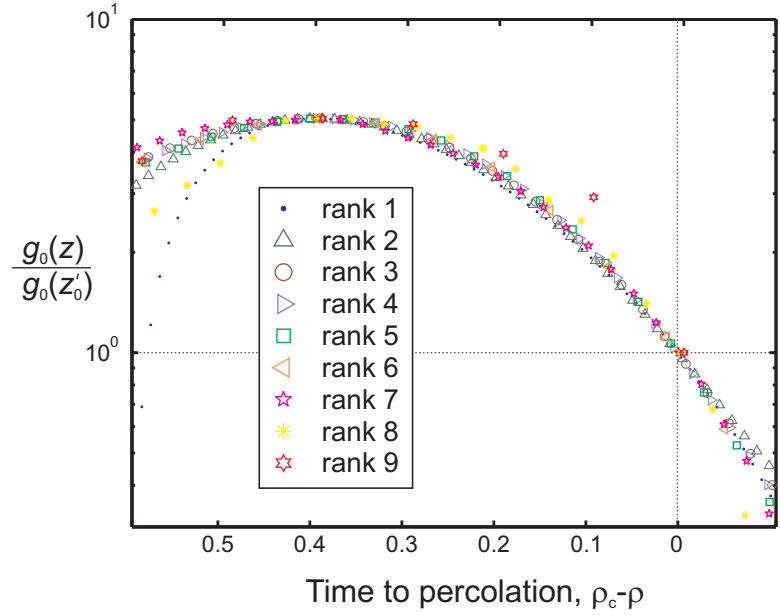}
\caption{Three-exponent scaling for
rank dynamics. The master Gaussians
$g_0(z)$ for different ranks collapse 
when using the renormalization given
by Eqs.~(\ref{r2exp}),(\ref{r3exp}).}
\label{fig_g0}
\end{figure}

The pure power and exponential laws in 
Figs.~\ref{fig_GRm}, \ref{fig_GRr} are just 
first-order approximations to the observed 
cluster distributions at percolation.
In both cases one sees the downward bending for 
small clusters and upward bending for large clusters.
These are not due to statistical fluctuations.
The downward bending for small clusters is explained by 
``deviations from scaling'' \cite{Hoshen+79}:
it can be shown analytically that the small clusters
do not yet obey the general scaling law of 
Eqs.~(\ref{GRm}), (\ref{GRr})
which holds only for large enough masses (ranks).
The upward bend at large clusters is due to finite-size
effects \cite{Hoshen+79,Margolina+84}: each large cluster 
that reaches outside the lattice boundary is ``seen''
as a number of smaller clusters, thus creating
the upward deviation from the pure power (exponential) 
law.
This phenomenon is especially important when the system is 
close to percolation and clusters of arbitrary large 
sizes have already been formed.
The appropriate scale corrections for the mass distribution
were studied by Hoshen et al. \cite{Hoshen+79} and Margolina
et al. \cite{Margolina+84}.

\begin{figure}[p]
\centering\includegraphics{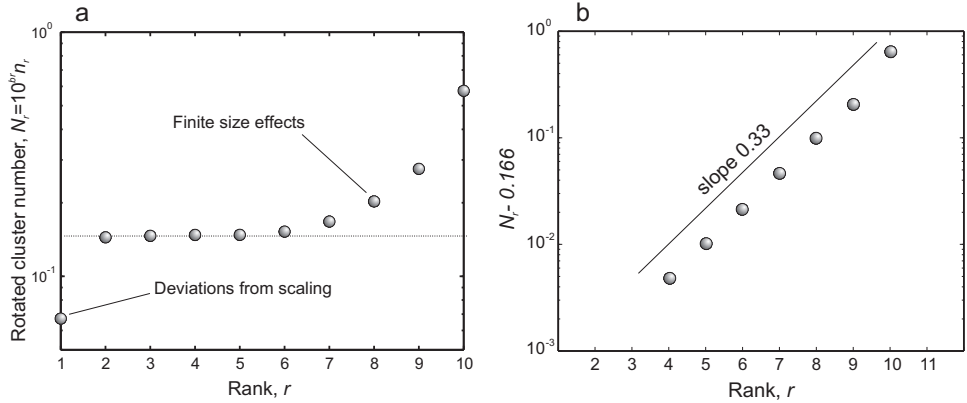}
\caption{Corrections to scaling.
The pure exponential rank distribution
of Eq.~(\ref{GRr}) suggests a horizontal
plateau for the normalized function 
$N_r=10^{br}n_r$, while the observed
values clearly deviate from the plateau
at small and large clusters (panel a).
The large cluster deviation is due to
finite size efffects and is described
by an exponential correction of
Eq.~(\ref{GRrc}) with $d\approx 0.33$
(panel b).}
\label{fig_Nr}
\end{figure}

To study the above phenomena it is convenient to consider 
the normalized functions
\[N_m:=m^{\tau-1}\sum_{m'\ge m} n_{m'},\qquad
N_r:=10^{br} n_r,\]
which, in the absence of scale corrections,
would become constants:
\[N_m=\frac{q_0}{\tau-1},\qquad
N_r=p_1\,c^{\tau-1}.\]
The function $N_r$ is shown in Fig.~\ref{fig_Nr}a;
it clearly deviates from the horizontal plateau
at both sides.

In case of the mass distribution, the corrections to
scaling are given by \cite{Margolina+84}:
\begin{equation}
\label{GRmc}
n_m(\rho_c)\simeq m^{-\tau}
\left(q_0+q_1\,m^{-\Omega}+q_L\,m^{1/D}L^{-1}\right),
\end{equation}
where $\Omega \approx 0.75$, $1/D=48/91$ is the 
universal mean 
cluster radius exponent, and $q_0,q_1,q_L$ are 
independent of $s$ and $L$.
The first additional term describes the deviation
from scaling for small clusters, while the second
one is responsible for finite-size effects.

For rank distribution, the ``deviations from scaling'' 
at lower clusters are only observed for $r=1$;
while the finite-size effects at large clusters are
clearly present for many ranks.
Accordingly, we propose the following correction
to scaling for the rank distribution:
\begin{equation}
\label{GRrc}
n_r(\rho_c)\simeq 10^{-br}
\left(p_0+p_L\,10^{d\,r}L^{-1}\right),~r>1. 
\end{equation}
with
\[d=\frac{1}{D}\,\log_{10}c \approx 0.33.\]
The observed value of $d$ can be estimated by plotting
$(n_r\,10^{br}-p_0)$ as a function of $r$ as shown 
in Fig.~\ref{fig_Nr}b.
The observed ranks $4\le r \le 9$ follow 
the predicted scaling (\ref{GRrc}) nicely.

Importantly, the corrections to scaling (\ref{GRrc})
act at all cluster sizes, 
so they can not be neglected even for the intermediate 
clusters, not only for the largest ones.
Indeed, their effect decreases with $L$, but this
decrease is very slow.
Notably, as shown by Morein et al. \cite{MNTG04}
(their Fig.~5)
even for lattices as large as $L=30,000$ during 
the process when clusters as large as 2\% of
the lattice size are removed, 
the cluster size distribution clearly exhibits 
the upward deviations at large ranks ($r=11,12,13$.)
For smaller systems these deviations become dominant
and may lead to an artificial decrease of the observed 
slope of cluster size distribution; this is demonstrated
in Fig.~\ref{fig_GRm},\ref{fig_GRr} and is also seen
in the analysis of Turcotte et al. \cite{Tur+99} 
(their Fig.~9). 

\subsubsection{Dynamics of scaling corrections}
\label{scalecorr}
Since the finite size effects play an important
role in shaping the observed cluster size 
distribution, it is worth studying their dynamics.
Specifically, we will be interested in transition
of the cluster size distribution from the convex 
shape (in semi- or bilogarithmic scale) at 
$\rho\ll\rho_c$ to formation of the upward bend
at percolation.

\begin{figure}[p]
\centering\includegraphics{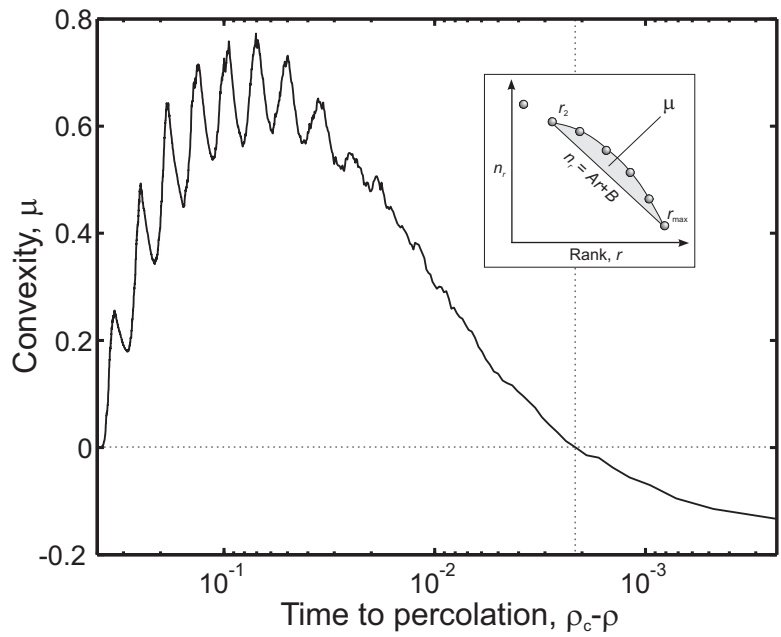}
\caption{Dynamics of scale corrections. 
A convexity measure $\mu(\rho)$ is 
defined by Eq.~(\ref{mconv}) and illustrated 
in the insert.
It is positive for convex, and negative for 
concave rank distribution.
The downward bend of the rank distribution
observed at early stages ($\mu>0$) is changed
to the upward one ($\mu<0$) for 
$(\rho_c-\rho)<2\cdot10^{-3}$.
See details in Sect.~\ref{scalecorr}.}
\label{fig_mu}
\end{figure}

For this we introduce a measure of convexity for
the rank distribution, defined as an area 
between $\log n_r(\rho)$ and a chord connecting its 
first and last points as shown in Fig.~\ref{fig_mu}
(the point $r=1$ is not considered being affected by
the deviations from scaling):
\begin{equation}
\label{mconv}
\mu(\rho):=\int_{2}^{r_{\rm max}} 
\left[\log_{10} n_r(\rho)-(A\,r+B)\right]
\,d\,r,
\end{equation}
with
\[A=\frac{\log_{10}\left(n_{r_{\rm max}}/n_2\right)}
{r_{\rm max}-2},
~B=\log_{10}\,n_2-2A.\]
The values of $\mu$ are positive when $n_r(\rho)$ 
is convex in semilogarithmic scale, negative when 
it is concave, and vanish when it is linear.
The measure $\mu(\rho)$ averaged over 1,000 runs on the 
lattice $L=2000$ is shown in Fig.~\ref{fig_mu};
the bell-shaped form of $\mu$ is decorated by 
the logperiodic oscillations for 
$(\rho_c-\rho)>10^{-2}$ explained by  
creation of new ranks, which temporarily 
increases convexity.
Zero level is crossed at about 
$(\rho_c-\rho)=2\cdot 10^{-3}$, after that
the rank distribution is concave.
A detailed analysis (not shown) demonstrates 
that the distribution is never exactly linear; 
the transition from convex to concave
shape is realized through the wave-shaped form
when the distribution is still convex for the 
lower $r$, but is already concave for the 
higher ones.
Qualitatively the same picture is observed
for the mass distribution $n_m(\rho)$
(in bilogarithmic scale).

The transformation
of the cluster size distribution prior to
percolation is not unlike a well-known pattern 
``upward bend'' first described by Narkunskaya and 
Shnirman \cite{NS90,NS94} in an early static model of 
defect development.
Later it was found in steel samples and seismicity 
of California \cite{RKB97},
and confirmed by the dynamical modeling of failure
in a hierarchical system (so-called colliding
cascade models) \cite{GKZN00,ZKG03}.

\subsection{Mass dynamics of a given rank}
\label{MDGR}
\begin{figure}[p]
\centering\includegraphics{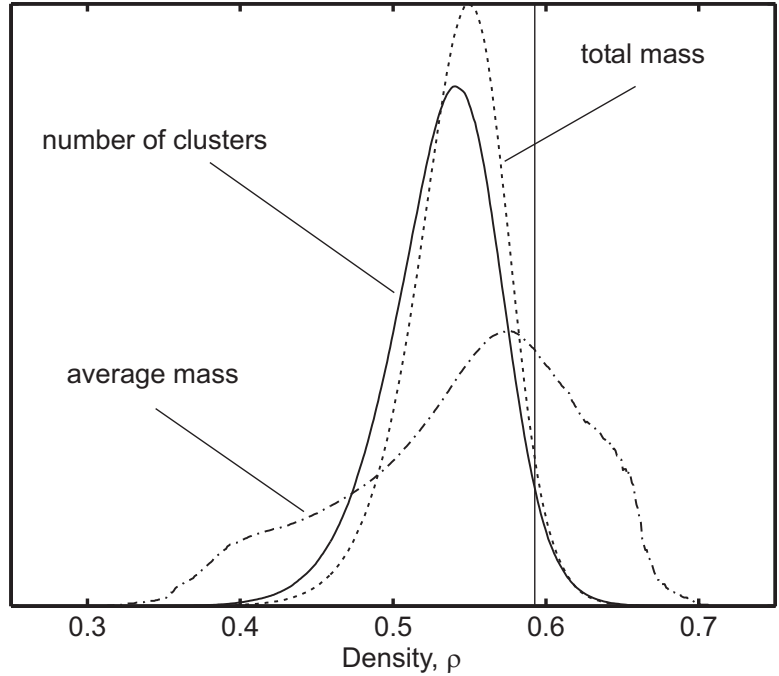}
\caption{Dynamics of number of clusters
$n_r$ (solid line), 
total mass $M_r$ (dashed line), 
and average mass $m_r$ (dash-dotted line)
for clusters of rank $r=5$.}
\label{fig_avem}
\end{figure}

Here we consider the dynamics of total and average
mass of rank $r$ clusters:
\begin{equation}
\label{mass}
M_r=\sum m\,n_{rm},\qquad
  m_r   =\frac{\sum m\,n_{rm}}{\sum n_r}
=\frac{M_r}{n_r}.
\end{equation}
Here $n_{rm}$ denotes the number of clusters of
rank $r$ and mass $m$.
Figure~\ref{fig_avem} shows $n_r$, $M_r$, and
$  m_r   $ for rank 5; the similar
picture is observed for other ranks.
It is tempting to use Gaussian approximation for 
$M_r$ and predict the Gaussian dynamics of 
$  m_r   $ (as a ratio of two
Gaussians) and relate their parameters.
Detailed analysis however demonstrates that under
this approach the peak of $  m_r   $ for 
ranks $r\ge 9$ should be observed after 
percolation; while in simulations this peak is 
always prior to percolation (not shown).
Note that one still might approximate 
$M_r$ and $  m_r   $ by Gaussians
with properly scaled parameters; such approximations
will be good for rough curve fitting, but
will fail to reproduce deeper properties of cluster
dynamics.
This demonstrates the general limitations of Gaussian 
approximations in the percolation problem.

\section{Cluster fractal structure}
\label{CFS}
\begin{figure}[p]
\centering\includegraphics{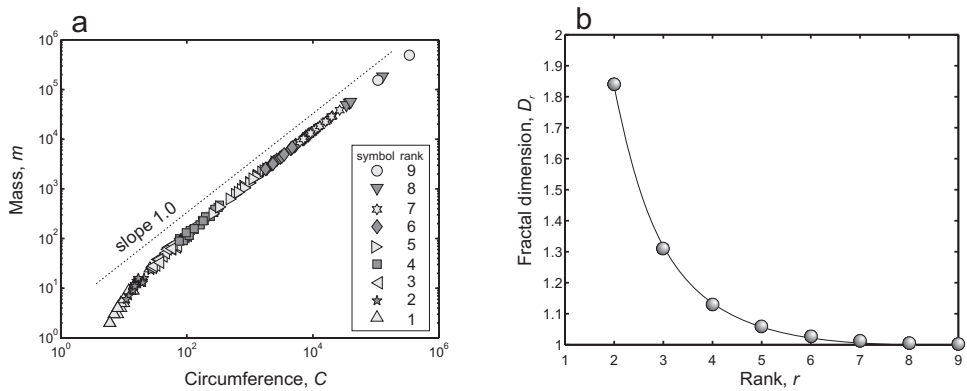}
\caption{Fractal structure of clusters.
a) Mass-circumference relation for
clusters of different ranks.
The asymptotic power relation with 
slope 1.0 is gradually develops as
rank increases.
b) Values of fractal dimension $D_r$
(Eq.~(\ref{circum})) for different ranks.
Both panels correspond to a 
$2,000\times2,000$ lattice at 
percolation.}
\label{fig_D}
\end{figure}

In this section we evaluate the fractal structure of 
clusters considering the mass-circumference 
relation
\begin{equation}
\label{circum}
m \propto C^{D_r},
\end{equation}
where $C$ is the number of empty neighbors of a 
cluster of mass $m$.
For percolation cluster at infinite grid we have $D_r=1$, 
which shows that the percolation cluster is a ``linear'' rather 
than a space-filling object \cite{SA}.
Figure~\ref{fig_D}a shows the cluster masses as a function
of their circumference for different ranks.
It is easily seen how the linear scaling 
$D_r=1$ gradually develops as rank increases.
Figure~\ref{fig_D}b shows the index $D_r$ estimated for 
$1\le r\le 9$.

\begin{figure}[p]
\centering\includegraphics{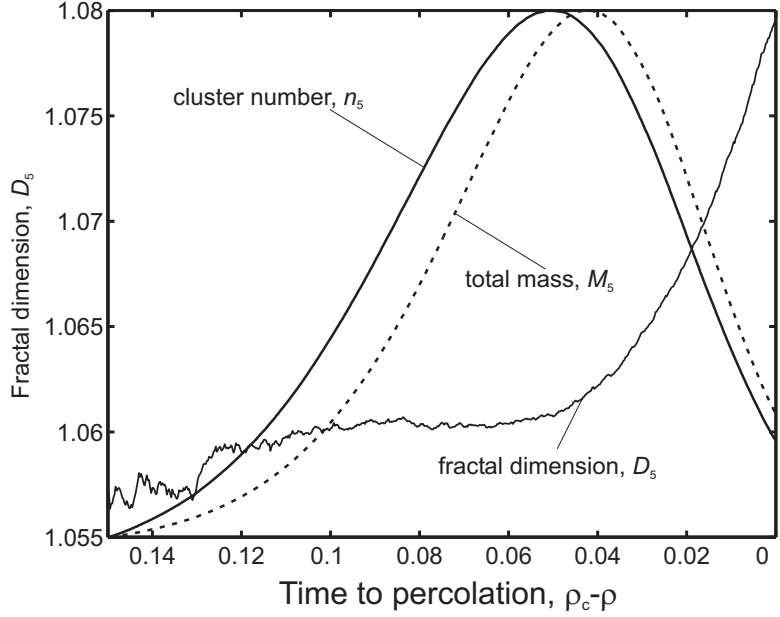}
\caption{Premonitory increase of cluster
fractal dimension.
The steady-state dynamics of fractal 
dimension $D_5$ (Eq.~\ref{circum}) changes,
and $D_5$ starts to increase, as
system approaches percolation.
Similar phenomenon is observed for
other ranks. }
\label{fig_fin}
\end{figure}

Figure~\ref{fig_fin} shows the dynamics of $D_5$
prior to percolation;
noteworthy, its steady state behavior is altered 
by a gradual increase as $\rho\to\rho_c$.
Similar increase is observed for clusters of other
ranks.

To explain the increase of $D_r$ we recall that the rate 
of cluster coalescence is proportional to their 
circumference (see e.g. \cite{GNT99}).  
Thus, for a given mass, clusters with a lower $D_r$ have
larger circumference, and a higher chance to coalesce.  
When a sufficient number of rank $r$ clusters have been
formed, the clusters with low $D_r$ start to coalesce 
leaving the high-$D_r$ clusters on the grid.  

Another reason for the increase of $D_r$ is 
the finite-size effects.
Specifically, this is an effect of having clusters 
that on an infinite grid have already gained higher 
ranks, but on our finite lattice are still small.

\section{Dynamical constraint}
\label{DC}
\begin{figure}[p]
\centering\includegraphics{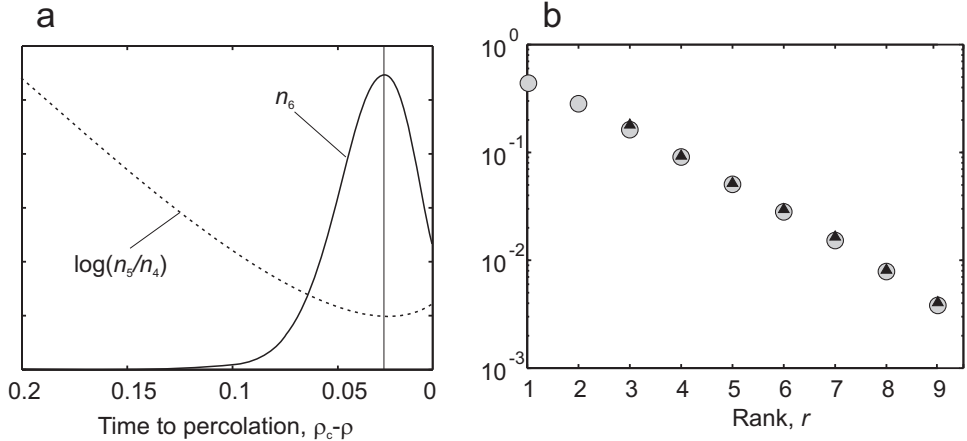}
\caption{Dynamical constraint for $n_r(\rho)$.
Dynamics of $\theta_4=\log(n_4/n_5)$ and 
$n_6$ is shown in panel a: peaks of two 
curves coincide.
The similar phenomenon is observed for
other ranks: panel b shows the times of 
maxima of $n_r$ (circles) and 
minima of $\theta_{r-2}$ (triangles)
for $3\le r\le 9$.}
\label{fig_peaks}
\end{figure}

Here we report an interesting regularity in
rank dynamics that put a notable constraint on
analytical modeling of percolation process.
Specificaly, we consider the slope 
between two consecutive points of the rank
distribution:
\[\theta_r(\rho):=\log\,\frac{n_r(\rho)}{n_{r+1}(\rho)}.\]
Dynamics of $\theta_4$ is shown in Fig.~\ref{fig_peaks}a
together with that of $n_6$.
Noteworthy, the peaks of two curves 
(minimum of $\theta_4$ and maximum of $n_6$)
coincide.
This happens to be true for all ranks:
positions of corresponding peaks are shown
as a function of rank in Fig.~\ref{fig_peaks}b.
Such perfect matching is very unlikely to be 
accidental.
Thus we conjecture that in order for $n_r(\rho)$ 
to properly describe the time-dependent 
behavior of rank population, 
the following system of differential equations 
must have a solution:
\begin{equation}
\label{constraint}
\left\{\begin{array}{l}
\dot{\theta}_r=0\\
\dot{n}_{r+2}=0
\end{array}
\right.
\Rightarrow
\left\{\begin{array}{l}
\dot{n_r}\,n_{r+1}-n_r\dot{n}_{r+1}=0\\
\dot{n}_{r+2}=0
\end{array}
\right.
\end{equation}

Applying this constraint to the three-exponent 
scaling of Eqs.~(\ref{r2exp}),(\ref{r3exp}),(\ref{g0})
we find
\begin{equation}
\label{sigmas}
\sigma_2=\sigma_1+\log_{10}\left(1-10^{-2\sigma_1}\right).
\end{equation}
According to (\ref{sigmas}), the observed value 
$\sigma_1=0.23$ gives
$\sigma_2=0.04$, which is 33\% larger than the 
observed value $\sigma_2=0.03$.
The discrepancy is due to the 
approximate character of the Gaussian approximation 
(\ref{g0}) for $g_0$. 

\section{Discussion} 
\label{discussion}
The goal of this study was to describe the evolution of
percolation model in terms of consecutive aggregation of 
smaller clusters into larger ones using the Horton-Strahler
hierarchical scheme.
First, this contributes to a novel understanding of
the percolation phenomenon as a time-dependent hierarchical
inverse cascade process.
Second, this allows one to test the validity of the
approach introduced by Gabrielov et al. \cite{GNT99} and 
further developed by Morein et al. \cite{MNTG04}
for a steady-state approximation to a general 
aggregation process.

We considered dynamical and scaling properties 
of site-percolation on a 2{\it D} square lattice.
Following \cite{GNT99} we described clusters by 
hierarchical trees that reflect the history of 
cluster formation;
the Horton-Strahler scheme was used to rank the
trees and thus the corresponding clusters.
We concentrated on the development of the first 
percolation cluster, thus working with a system that 
does not exhibit the steady-state dynamics,
contrary to the studies \cite{GNT99,MNTG04} 
that have developed mean-field steady-state 
approximations to the system.

Combining the results obtained in the classical percolation 
studies with the Tokunaga constraint on the cluster branching 
structure we derived various rank-dependent scaling 
laws connecting the number $n_r$ of clusters of rank $r$,
their average mass $m_r$, and the rank $r$. 
We have compared the parameters of these laws with
those predicted and observed by \cite{GNT99,MNTG04} in 
steady-state aggregation models.
The values of parameters are shown to be in a perfect 
agreement, confirming the validity of the approach used 
in \cite{GNT99,MNTG04}.
In absence of the steady-state behavior,
we derived the time-dependent versions of the scaling laws.
We reported the three-index scaling (\ref{r2exp}),
(\ref{r3exp}) for the number $n_r(\rho)$ of clusters of 
a given rank, which deviates from the classical two-exponent 
scaling for masses.

We studied in detail the transition of the system
from earlier stages to the vicinity of percolation
and reported several characteristic phenomena 
observed as $\rho\to\rho_c$.
They include transformation of the cluster size 
distribution
not unlike that observed in seismicity, steel samples,
and previous models of hierarchical fractures
\cite{NS90,RKB97,GKZN00,ZKG03}; and
increase of the cluster fractal dimension.
In our simple model these phenomena are partly explained
(qualitatively as well as quantitatively) by finite-size 
effects; nevertheless we believe that they should
not be neglected as irrelevant side-effects of
numerical simulation.
In fact, in practice we often work with systems that
are described by intermediate depth hierarchies
(in other words they have intermediate number of 
degrees of freedom).
The percolation results related to small and
intermediate lattices might be of high relevance in
describing such systems.
In addition, simulations on large
lattices ($L=30,000$) performed by 
Morein et al. \cite{MNTG04}
show that finite-size effects are still present 
even for large $L$. 

We have formulated  
an empirical constraint of Eq.~(\ref{constraint}) 
for the time-dependent behavior of rank
population size $n_r(\rho)$; 
the constraint follows very clearly from the observed 
values of $n_r(\rho)$.
It would be interesting to check this condition 
in real systems traditionally described by the 
percolation model. 

Our closing remark is on the index $\tau$ of 
cluster mass distribution at percolation
(Eq.~(\ref{GRm})).
The studies of Gabrielov et al. \cite{GNT99} and 
Morein et al. \cite{MNTG04} predict $\tau=2$;
which slightly deviates from the well established 
theoretical value of the Fisher exponent 
$\tau=187/91\approx 2.05$. 
The index of the mass distribution is an essential
characteristic of a system, thus even this slight
difference of 2.5\% might seem disappointing
implying the intrinsically approximate character 
of the modeling of \cite{GNT99,MNTG04}.
In fact, this implication is not true.
To validate the approach of \cite{GNT99,MNTG04} we 
notice that the Fisher exponent is tightly
connected to the precise count of cluster 
particles on a site-level, hardly feasible in practice.
At the same time, the studies \cite{ZLK99,CZ03} 
have demonstrated that when we ``characterize the size
distribution of clusters in a way that circumvents
the site-level description'' considering any
``macroscopic measure of the length scale of the
cluster'', the exponent of the corresponding 
scaling law becomes $2$, universally for all 
2{\it D} systems.
An example of a ``macroscopic measure'' is
the linear size in arbitrary direction, 
the radius of gyration, 
the diameter of the covering disk, etc.
Clearly, the modeling of \cite{GNT99,MNTG04}
deals with such a macroscopic measure of 
cluster size, and hence predicts the correct 
scaling exponent.

\bigskip
{\bf Acknowledgment.}
We are sincerely grateful to Bill Newman 
for numerous focused discussions; his critics and advice 
have helped significantly in organizing and presenting 
the results.
We thank Vladimir Keilis-Borok, Gleb Morein, and 
Donald Turcotte for their continuous interest to the work 
and David Shatto for help in preparing the manuscript.
This study was partly supported by NSF, grant 
ATM 0327558.

\end{document}